\documentstyle[preprint,aps]{revtex}

\begin{document}

\draft

\title{Mixed-State Quasiparticle Spectrum for d-wave Superconductors}

\author{Yong Wang and A. H. MacDonald}

\address{Department of Physics, Indiana University,
Bloomington, IN 47405}

\date{\today}

\preprint{IUCM95-015}

\maketitle

\begin{abstract}

Controversy concerning the pairing symmetry of
high-$T_c$ materials has motivated an interest in
those measureable properties of superconductors for which
qualitative differences exist between the s-wave and d-wave cases.
We report on a comparison between the microscopic electronic
properties of d-wave and s-wave superconductors in the mixed state.
Our study is based on self-consistent numerical solutions of the
mean-field Bogoliubov-de Gennes equations for phenomenological BCS
models which have s-wave and d-wave condensates in the absence
of a magnetic field.  We discuss differences
between the s-wave and the d-wave local density-of-states, both near
and away from vortex cores. Experimental implications for both
scanning-tunneling-microscopy measurements and specific heat
measurements are discussed.

\end{abstract}

\pacs{\\ PACS: 74.60.Ec, 74.72.-h}

Since shortly after the discovery of high-temperature superconductors
(HTSC), there has been great interest in determining the pairing
symmetry of the
order parameter\cite{schrieffer}. In the absence of
disorder, low temperature electronic properties in the Meissner
state of a d-wave superconductor
differ qualitatively from those of a conventional s-wave
superconconductor because
of the existence of nodes in the gap function.  These
differences can in principle be used to identify the pairing
symmetry, although
strong anisotropy and the complicated nature of the materials
have conspired to make conclusive
experiments difficult.  (Recent work is strongly suggestive of
$d_{x^2-y^2}$ pairing.)  It is also of interest to study differences
between the mixed-states of d-wave and s-wave type II superconductors.
In the mixed-state magnetic flux will
penetrate the superconductor and form an Abrikosov vortex lattice.
Low-lying quasiparticle excitations will then exist for both pairings,
although in the conventional case they must be bound to the
vortex core where the
order parameter vanishes.  The existence of bound quasiparticle
states was first
predicted by Caroli, de Gennes and Matricon\cite{caroli} when
they studied an
isolated vortex line in a conventional superconductor.
Experimentally, these
quasiparticles have been observed in scanning-tunneling-microscopy
(STM) measurements\cite{Hess}.
For the d-wave case, progress has recently been made on both
experimental and
theoretical fronts.  Volovik has used semiclassical
approximations\cite{volovik,maki} to
calculate the density-of-states (DOS) at the Fermi energy $N (0)$
for the mixed state of a $d_{x^2-y^2}$ superconductor in a weak
magnetic field $H \ll H_{c2}$.  He found a finite $N(0)$ in the
absence of disorder proportional
to $H^{1/2}$ compared to the $H^{1}$ behavior expected for
 conventional superconductors in
the same approximation.  This prediction appears to be
in accord with recent
measurements of the magnetic field dependence of the low-temperature
specific heat\cite{speci_heat,spec_review} in high $T_c$ materials.
However, the short coherence length of high $T_c$ materials
raises some uncertainty about the detailed applicability of a
semi-classical analysis
and motivates a fully microscopic study of the same problem.
In this Rapid Communication we report on such a study.

Application of microscopic mean-field-theory to inhomogeneous
states of superconductors gives
rise to the Bogoliubov-de Gennes (BdG) equations\cite{deGennes}.
Motivated by the STM experiments of Hess\cite{Hess}, numerical
solutions of the
BdG equations have been obtained for both
continuum\cite{gygi,dorsey} and
lattice\cite{cincinatti} models of a superconductor containing an
isolated vortex. This work has recently been generalized to the
 case of isolated
vortex in a d-wave superconductor\cite{mcmaster}.  According to
Volovik, the DOS in the
mixed state of a d-wave superconductor is dependent on the typical
distance between vortices so that for the present study
it is necessary to
solve the BdG equations for the vortex-lattice state of
a d-wave superconductor.

To model decoupled $CuO_{2}$ layers we consider
single-band Hamiltonians on a 2D square lattice
with nearest neighbour
hopping and both on-site and nearest-neighbor interactions:
\begin{mathletters}
\begin{equation}
H = H^0 + H^{'},
\label{mlett:1}
\end{equation}
\begin{equation}
H^0  = - \sum_{<ij> \sigma} (t_{ij}c^{\dag}_{j \sigma} c_{i \sigma}
 + t_{ji}c^{\dag}_{i \sigma}c_{j \sigma})
-\sum_{i \sigma}\mu \hat n_{i \sigma},
\label{mlett:2}
\end{equation}
\begin{equation}
H^{'} = U \sum_i \hat n_{i \uparrow} \hat n_{i \downarrow}
+ {V \over 2} \sum_
{<ij> \sigma
\sigma'} \hat n_{i \sigma} \hat n_{j \sigma'}.
\label{mlett:3}
\end{equation}
\end{mathletters}
Here $i$ and $j$ are site labels, the angle brackets
in Eq.(~\ref{mlett:1})
imply the restriction to neighboring sites,
$\hat n_{i \sigma} = c^{\dag}_{i \sigma} c_{i \sigma}$
is the electron number operator on site $i$,
 and $\mu$ is the chemical potential.
We will assume that the screened magnetic field inside
the superconductor can
be taken to be constant; for high $T_c$ materials this is
a good approximation
except for external fields extremely close to $H_{c1}$.  In
a one-band lattice model
the magnetic field appears in the hopping amplitudes:
\begin{equation}
t_{ij} =  t e^{-i {e \over {\hbar c}}
\int_{\vec r_i}^{\vec r_j}d \vec r \cdot
\vec A (\vec r)}
\label{tij}
\end{equation}
where $t$ is the nearest neighbor hopping amplitude in zero field and
 $\nabla \times \vec A( \vec r) = \vec h (\vec r)$.  We
report results below for
two different models.  The `s-wave' model has on-site attraction
$U < 0$, and no nearest-neighbor interaction.
For the `d-wave' model, we set $U>0$ and $V<0$. When the magnetic
field is set
to zero the mean-field BCS gap equations are readily solved
for either model by
using translational invariance on the lattice.  For the `s-wave' case
the pairing
self-energy in the ordered state is proportional
to the order parameter
\begin{equation}
\Phi^{s}_{i}=<c^{\dag}_{i\uparrow} c^{\dag}_{i \downarrow}>,
\label{sw}
\end{equation}
while for the `d-wave' model it is proportional to
\begin{equation}
\Phi^{d}_{i}= {1 \over 4}
\sum_{\delta}(-)^{\delta_y}<c^{\dag}_{i\uparrow}
 c^{\dag}_{i+\delta \downarrow}>,
\label{dw}
\end{equation}
where the sum is over the nearest-neighbors of site $i$,
represented by the unit vectors
$\hat \delta = \pm {\hat e}_x, \pm {\hat e}_y$.
(In the homogeneous zero-field
states $\Phi^{s}_{i}$ and $\Phi^{d}_{i}$ are independent of $i$.)
For both models
the numerical values of the interaction parameters have been
chosen to give a
zero-temperature coherence length
(estimated from the pair wavefunction) $\sim 4 a$,
as in high $T_c$ materials.  The results reported below
were calculated for the case of a band filling factor $<n>=0.8$.
For the `s-wave' model, we set $U=-3.5t$ and $V=0.0$
 while for the d-wave model we choose $U=2.1t$,
and $V=-2.1t$.  For layer separations and bandwidths appropriate for
models of high-temperature superconductors,
the penetration depths (at $T=0$) for these models are
several hundred times larger than the conherence lengths so that the
models do indeed descripe strongly type-II superconductors.

To study vortex lattice states, we introduce magnetic unit cells
, each containing two superconducting flux quanta:
$\Phi_0=hc/2e$. The size of a unit cell is $N_x a \times N_y a$ in
general, where $a$ is the lattice constant.
We then define magnetic Bloch states labeled by a magnetic wave
vector $\vec k$, a site index $i$ within the magnetic unit cell,
and a spin index $\sigma$ and
denote the corresponding creation and annihilation operators by
$c^{\dag}_{\vec k i \sigma}$ and $c_{\vec k i \sigma}$.
In mean-field theory these are related to the quasiparticle
creation and annihilation
operators by
\begin{mathletters}
\begin{equation}
c^{\dag}_{\vec k i \uparrow} = \sum_{\alpha}u^{\alpha}_{i}(\vec k)
\gamma^{\dag}_{\vec k \alpha \uparrow}
-\sum_{\alpha}v^{\alpha \ast}_{i}(\vec k)
\gamma_{\vec k \alpha \downarrow},
\label{mlett:B1}
\end{equation}
\begin{equation}
c_{-\vec k i \downarrow} =
\sum_{\alpha}v^{\alpha}_{i}(\vec k) \gamma^{\dag}_{\vec k
 \alpha \uparrow} + \sum_{\alpha}u^{\alpha \ast}_{i}(\vec k)
\gamma_{\vec k
\alpha \downarrow},
\label{mlett:B2}
\end{equation}
\end{mathletters}
where $u^{\alpha}(\vec k)$ and $v^{\alpha}(\vec k)$ are
determined by solving the
BdG equations:
\begin{equation}
\left(\begin{array}{cc}
H^0(\vec k)&UF_{1}(\vec k)+VF_{2}(\vec k)\\
UF_{1}^{\ast}(\vec k)+VF_{2}^{\ast}(\vec k)&-H^{0 \ast}(\vec k)
\end{array}\right) \left(\begin{array}{c}
u^{\alpha}(\vec k)\\
v^{\alpha}(\vec k)
\end{array}\right) = E^{\alpha}(\vec k) \left(\begin{array}{c}
u^{\alpha}(\vec k)\\
v^{\alpha}(\vec k)
\end{array}\right),
\label{bdg}
\end{equation}
where $H^0$, $F_1$, and $F_2$ are  $N_xN_y \times N_xN_y$ matrices.
The off-diagonal blocks in this matrices are the pairing
self-energies and these
can be expressed in terms of quasiparticle amplititudes.
The on-site interaction contribution is diagonal
in site indices with
\begin{equation}
(F_{1}(\vec k))_{ii}\equiv <c^{\dag}_{\vec k i\uparrow}
c^{\dag}_{-\vec ki \downarrow}>=
-\sum_{\alpha} \tanh{{\beta E^{\alpha}(\vec k)}
\over {2}}u^{\alpha}_{i}(\vec k)v^{\alpha \ast}_{i}(\vec k),
\label{pot1}
\end{equation}
while the nearest neighbor interaction contribution is
\begin{equation}
(F_{2}(\vec k))_{ij}\equiv <c^{\dag}_{\vec k i\uparrow}
c^{\dag}_{-\vec kj\downarrow}>\Delta_{j,i+\delta}
=-{1 \over 2}\sum_{\alpha}
\tanh{{\beta E^{\alpha}(\vec k)} \over {2}}(u^{\alpha}_{i}(\vec k)
v^{\alpha \ast}_{j}(\vec k)
+ u^{\alpha}_{j}(\vec k)v^{\alpha \ast}_{i}(\vec k
))\Delta_{j,i+\delta}.
\label{pot2}
\end{equation}
Here $\beta = 1/k_BT$ and
$$\Delta_{j,i+\delta}=\cases{\delta_{j,i+\delta},
&if site $i+\delta$ and site $i$ are in the same cell;
 \cr \delta_{j,i-N_\delta\delta},&if site $i+\delta$ and
site $i$ are in different cells, \cr} $$
where $N_\delta$ is the number of sites
along $\delta$ direction in a cell.
Eqs.\ (\ref{bdg}),\ (\ref{pot1}) and (\ref{pot2})
constitute a set of
self-consistent equations, whose solutions can be obtained
numerically by
iteration.

Typical\cite{latticestructure} self-consistent
results for the order parameter
of the d-wave model in the vortex lattice state at $T=0$ are
shown in Fig.~\ref{order}.
 The size of the cell is $28a \times 56a$,
corresponding to a field of $H=\Phi_0/(28a)^2$; if we
associate $a$ with the
typical Cu-Cu distance in high $T_c$ materials this corresponds
to a magnetic field
$\sim 10 {\rm Tesla}$.   The value of the order parameter at
the mid-point between
neighboring vortices is $\Phi^d_{i,max}(H)=0.063$, which may be
compared with the
value $\Phi^{d} = 0.065$ obtained at $T=0$ for the Meissner state
of the same model.
 The magnetic field
suppresses superconductivity everywhere in the system; numerical
calculations at stronger
fields are consistent\cite{sfsc} with upper critical
fields $H_{c2} \sim 100 {\rm Tesla}$
for the model we study, as expected from the zero-temperature
coherence length.
We remark that the extended s-wave component of the order
paramater, permitted
by symmetry \cite{volovik,mcmaster,houston} in the
vortex lattice state and
possibly of experimental relevance\cite{mcmaster2,neutrons}
in high $T_c$ materials,
is smaller than the d-wave component by about
two orders of magnitude for
the model parameters we have chosen.

We define the local density-of-states (LDOS) on site $i$ by
\begin{equation}
N_i(E)=-{{1} \over {N_c}}\sum_{\vec k,\alpha}
[|u_i^{\alpha}(\vec k)|^2f'(E^{\alpha}(\vec k)-E)
+|v_i^{\alpha}(\vec k)|^2f'(E^{\alpha}(\vec k)+E)],
\label{dos_def}
\end{equation}
where $f(E)=(\exp(\beta E) +1)^{-1}$, and $N_c$ is the
number of the magnetic
cells in the system.  $N_i(E)$ is
proportional to the differential tunneling
conductance\cite{gygi,tinkham}
which is measured in an STM experiment.

We show in Fig.~\ref{dos_fig1} $N_i(E)$ at a site
midway between two neighboring vortices
for both s-wave and d-wave superconductors
at zero field and at several different
finite field strengths.  In a field, the density of states
at the Fermi level,
$N_i(0)$, is much larger for the d-wave case than
for the s-wave case as
predicted by Volovik.  (We are unable to solve the BdG equations
at weak enough fields to verify the
expected $H^{1/2}$ behavior, although
as seen in the inset
our results are consistent with this prediction.)
It is presumably this density-of-states away from
the d-wave vortex cores which is responsible for the enhancement of
the low-temperature
specific heat of high $T_c$ superconductors in a field seen by Moler
{\it et al.} \cite{speci_heat,detail}.  In the s-wave
case, the zero-field gap remains quite well defined out to
fairly strong
magnetic field strengths although, in contrast with the
semiclassical result, the
density of states is not strictly zero at any energy.  The
size of the gap decreases
with increasing field as expected.  The sharp peaks in zero
field DOS are of different
origins. The peak closest to the Fermi energy is due to
superconductivity while
the second peak reflects the Van Hove singularity in the
band structure. These
peaks are smeared out in finite field.  In Fig.~\ref{dos_fig2}
we show $N_i(E)$ for
a site at the center of a vortex core.
In both s-wave and d-wave cases, we find large peaks near
the Fermi energy,
reflecting resonances which will evolve into quasiparticle
bound states in the
limit of isolated vortices. The
positions of the peaks are distinctly different
in two cases. In the d-wave
model, the LDOS peak is not as strong and is clearly centered at
the Fermi energy\cite{remark}. In the s-wave case
two quasiparticle LDOS peaks are visible and the lowest
energy of these is
clearly located away from the Fermi energy. The scale of
the separation between s-wave
bound-state peaks is $\sim 0.2 t$, in accord with expectations
based on the size
of the gap and the band width.  For these short coherence
length models, the
separation between quasiparticle bound states is large
enough to be resolved
so that low-temperature STM experiments would see a double-peak
structure in high $T_c$ materials if they had s-wave symmetry,
rather than the zero-bias peak observed\cite{Hess} in
conventional superconductors.

This work was supported by the National Science Foundation under
grant DMR-9416906.  The authors are grateful to
D. Arovas, A.J. Berlinsky,
C. Kallin, and M. Norman for helpful conversations.

\begin{figure}
\caption{The amplitude of a d-wave order
parameter $\Phi^d_i$ (z axis)
in a unit cell for the square lattice solution at $T=0$.
The size of the cell is $28a \times 56a$,
corresponding to a field $H=\Phi_0/(28a)^2$.}
\label{order}
\end{figure}

\begin{figure}
\caption{Quasiparticle local density-of-states profiles
at $T = 0$
away from vortex cores
for different magnetic fields are plotted in (a) for the d-wave
model and in  (b) for the s-wave model. The largest cell size
corresponds to the weakest magnetic field. The zero field
limits for the same models are also plotted. All
energies are in units of t
and measured from the Fermi energy. The inset is
the spatially averaged DOS
$N(0)$ vs $H^{1/2}$.}
\label{dos_fig1}
\end{figure}

\begin{figure}
\caption{Quasiparticle local density of states profiles
at $T=0$ at the center of the vortex cores.
Energies are in units of t and measured from
the Fermi energy.}
\label{dos_fig2}
\end{figure}


\begin{references}
\bibitem{latticestructure} We have obtained solutions for
both square
and triangular vortex lattices.  For
the models considered the triangular
lattice always has lower energy.   However, the dependence
of the quasiparticle density-of-states
on the vortex lattice structure is not of importance
for the issues addressed here
and all the results reported here were calculated
for the square vortex
lattice state.
\bibitem{remark} Bound states are localized in the core area with
$E < \Delta_\infty$. The maximum of the LDOS $N_i(E)$ shifts
to the Fermi
energy from higher energies as the center of vortex core
is approached.
\end{references}
\end{document}